\documentclass[amsmath,amssymb,aps,prb,superscriptaddress,longbibliography,citeautoscript]{revtex4-1}
\usepackage[margin=1in]{geometry}			
\usepackage{soul,xcolor}    
\setstcolor{blue}           
\usepackage{graphicx,epsfig}
\usepackage{xcolor}
\usepackage{hyperref}
\hypersetup{
    colorlinks = true,
}
\usepackage{comment}

\begin{document}
\author{Ming-Guang Hu}
 \thanks{These authors contributed equally.}
\affiliation{Department of Chemistry and Chemical Biology, Harvard University, Cambridge, Massachusetts, 02138, USA.}
\affiliation{Department of Physics, Harvard University, Cambridge, Massachusetts, 02138, USA.}
\affiliation{Harvard-MIT Center for Ultracold Atoms, Cambridge, Massachusetts, 02138, USA.}

\author{Yu~Liu}
 \thanks{These authors contributed equally.}
\affiliation{Department of Physics, Harvard University, Cambridge, Massachusetts, 02138, USA.}
\affiliation{Department of Chemistry and Chemical Biology, Harvard University, Cambridge, Massachusetts, 02138, USA.}
\affiliation{Harvard-MIT Center for Ultracold Atoms, Cambridge, Massachusetts, 02138, USA.}

\author{Matthew A. Nichols}
\affiliation{Department of Chemistry and Chemical Biology, Harvard University, Cambridge, Massachusetts, 02138, USA.}
\affiliation{Department of Physics, Harvard University, Cambridge, Massachusetts, 02138, USA.}
\affiliation{Harvard-MIT Center for Ultracold Atoms, Cambridge, Massachusetts, 02138, USA.}

\author{Lingbang~Zhu}
\affiliation{Department of Chemistry and Chemical Biology, Harvard University, Cambridge, Massachusetts, 02138, USA.}
\affiliation{Department of Physics, Harvard University, Cambridge, Massachusetts, 02138, USA.}
\affiliation{Harvard-MIT Center for Ultracold Atoms, Cambridge, Massachusetts, 02138, USA.}
\author{Goulven Qu\'{e}m\'{e}ner}
 \affiliation{Universit{\'e} Paris-Saclay, CNRS, Laboratoire Aim{\'e} Cotton, 91405, Orsay, France}

 \author{Olivier Dulieu}
 \affiliation{Universit{\'e} Paris-Saclay, CNRS, Laboratoire Aim{\'e} Cotton, 91405, Orsay, France}

\author{Kang-Kuen Ni}
\affiliation{Department of Chemistry and Chemical Biology, Harvard University, Cambridge, Massachusetts, 02138, USA.}
\affiliation{Department of Physics, Harvard University, Cambridge, Massachusetts, 02138, USA.}
\affiliation{Harvard-MIT Center for Ultracold Atoms, Cambridge, Massachusetts, 02138, USA.}

\title{{\Large State-to-state control of ultracold molecular reactions}}

\date{\today}
\begin{abstract}
\begin{large}

\textbf{Quantum control of reactive systems has enabled microscopic probes of underlying interaction potentials~\cite{klein2017directly,de2020imaging}, the opening of novel reaction pathways~\cite{gordon2018quantum,puri2017synthesis}, and the alteration of reaction rates using quantum statistics~\cite{ospelkaus2010quantum}. However, extending such control to the quantum states of reaction outcomes remains challenging. In this work, we realize this goal through the nuclear spin degree of freedom, a result which relies on the conservation of nuclear spins throughout the reaction~\cite{quack1977detailed,oka2004nuclear}. Using resonance-enhanced multiphoton ionization spectroscopy to investigate the products formed in bimolecular reactions between ultracold KRb molecules, we find that the system retains a near-perfect memory of the reactants\textquoteright\ nuclear spins, manifested as a strong parity preference for the rotational states of the products. We leverage this effect to alter the occupation of these product states by changing the coherent superposition of initial nuclear spin states with an external magnetic field. In this way, we are able to control both the inputs and outputs of a bimolecular reaction with quantum state resolution. The techniques demonstrated here open up the possibilities to study quantum interference between reaction pathways~\cite{brumer2000identical}, quantum entanglement between reaction products~\cite{gong2003entanglement}, and ultracold reaction dynamics at the state-to-state level~\cite{nesbitt2012toward}. }

\end{large}

\end{abstract}
\maketitle
\large

Controlling reactions between molecules at the level of individual quantum states is a long-standing goal in chemistry and physics 
that has prompted the development of a variety of highly-refined techniques for the state-selective preparation of reactants~\cite{chang2015spatially,vitanov2017stimulated,shapiro2002coherent} and examination of products~\cite{lin2003state,ashfold2006imaging}. Such experimental methods have pushed the boundaries of our understanding about the complex chemical interactions that govern the dynamics of reactive systems~\cite{yang2007state}. Recent advances in cold and ultracold techniques~\cite{balakrishnan2016perspective,toscano2020cold}
have furthered this frontier by allowing for control over all quantum degrees of freedom of molecules, including electronic, rotation, vibration, nuclear spin, and partial waves of scattering~\cite{quemener2012ultracold,jankunas2015cold}. These capabilities have enabled studies of scattering resonances which probe interaction potentials with exceptional resolution~\cite{klein2017directly,de2020imaging}, steering of reactive pathways to form selected product species~\cite{gordon2018quantum,puri2017synthesis}, and modification of long-range interactions to dramatically alter reaction rates~\cite{ospelkaus2010quantum,ni2010dipolar,hall2012millikelvin,perreault2017quantum,guo2018dipolar}. While such results demonstrate the power to efficiently change the overall reactivity through the quantum states of reactants, control over the quantum states in which products are formed has yet to be realized~\cite{bohn2017cold}. Achieving this state-to-state control could facilitate studies of quantum interference~\cite{brumer2000identical} and quantum entanglement~\cite{gong2003entanglement} in chemical reactions. Despite much experimental effort, however, even the ability to detect, let alone manipulate, the quantum states in which product species emerge in these cold and ultracold reactions has remained challenging.

Quantum control over different aspects of reaction outcomes have previously been achieved through coherent manipulation of atoms and molecules via ultrafast optical pulses~\cite{shapiro2002coherent,brif2010control}, as well as through excitation of molecular vibrational states~\cite{crim1996bond}. These techniques have successfully been applied to systems which exhibit direct reactions.
Low-temperature reactions, on the other hand, are typically barrierless, and proceed through the formation of intermediate complexes~\cite{bell2009ultracold}, which can obstruct the possibility for product control in these systems.
These complexes live for a duration that is many times longer than the characteristic rovibrational timescale, leading to a mixing of the energy of the complex into its various degrees of freedom. Such an effect is expected to result in a statistical distribution of the reaction products in which the effect of the reactant state preparation is strongly suppressed~\cite{levine2009molecular}. One notable exception to this behavior, however, is that spins of the nuclei involved tend to remain unchanged throughout reactions~\cite{quack1977detailed,oka2004nuclear}. This phenomenon has been observed in both complex-forming~\cite{uy1997observation,cordonnier2000selection} and direct~\cite{fushitani2002nuclear,momose2005chemical} bimolecular reactions, as well as in photodissociations\cite{schramm1983nuclear,webb2007imaging}
, and was attributed to weak coupling between nuclear spins and other degrees of freedom. Nuclear spins therefore provide a promising handle via which control over ultracold reactions can be exerted.

Previous works that established nuclear spin conservation in chemical reactions focused on hydrogen-containing species, due largely to their relevance in astrochemical processes. 
In reactions that feature higher spin nuclei and longer-lived intermediate complexes, such as those encountered in the ultracold regime~\cite{mayle2013scattering,christianen2019quasiclassical}, the fate of nuclear spins is less certain and remains to be experimentally examined. Prior studies typically carried out such examinations in systems involving cascaded reactions, using reactants with statistically mixed nuclear spins. Ultracold molecules, in contrast, can be prepared in single nuclear spin states, and react via single collisions. Hence, they provide a unique platform to cleanly probe the behavior of nuclear spins in chemical reactions.
Furthermore, the clean and highly-tunable nature of these systems, combined with their potential for high-resolution spectroscopy, provides access to new opportunities for the manipulation of chemical processes, including state-to-state control of reactions.

\begin{figure}
\centering
\includegraphics[width=6.3 in]{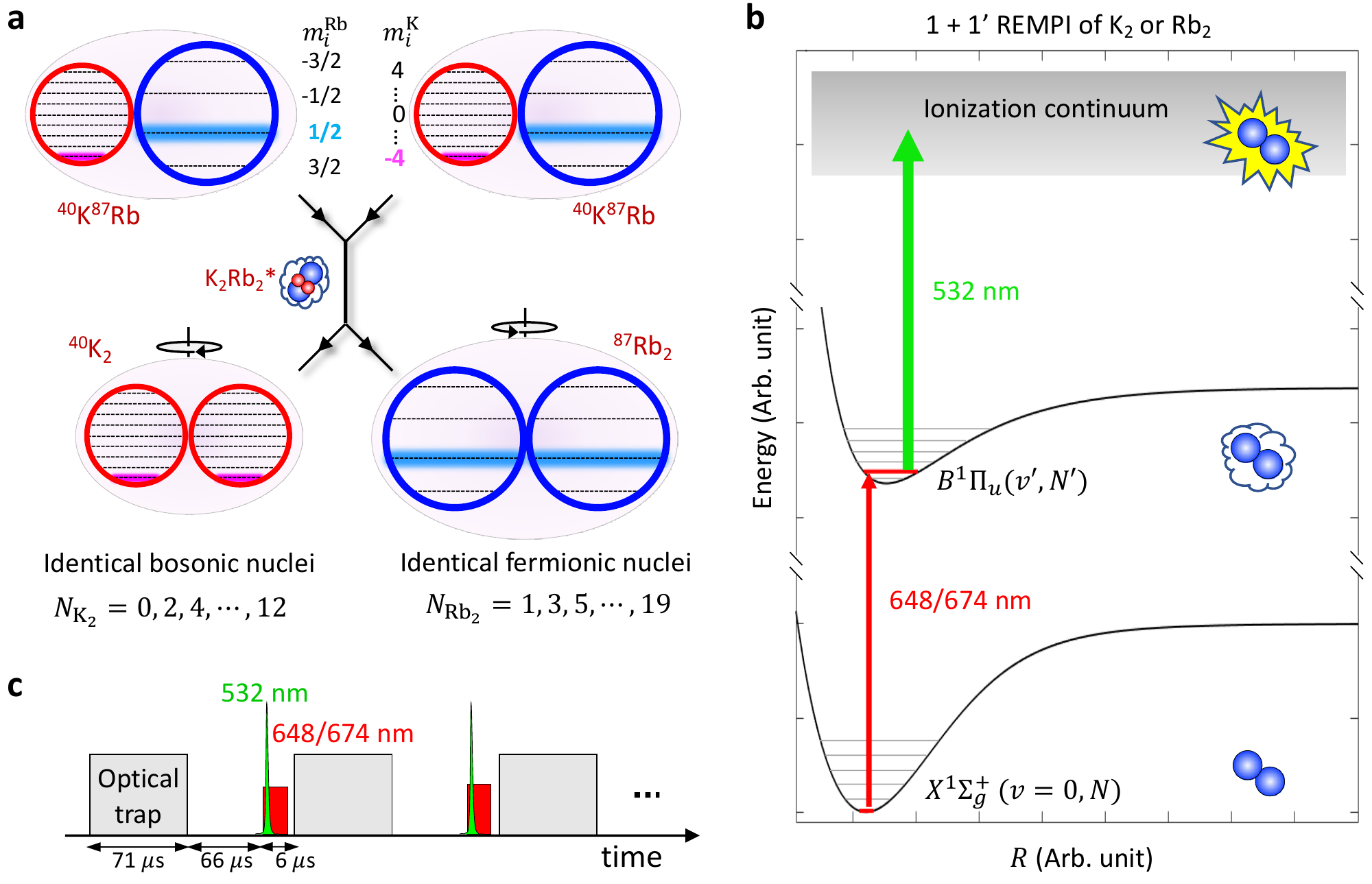}
\caption{\textbf{Probing the behavior of nuclear spins in reactions between KRb molecules.}
\textbf{a}, Schematic illustration of the conservation of nuclear spin throughout the reaction KRb + KRb $\rightarrow$ K$_2$Rb$_2^*$ $\rightarrow$ K$_2$ + Rb$_2$ at $B=30$ G. In this case, the initial reactant nuclear spin states $m_i^\text{K}=-4$ and $m_i^\text{Rb}=1/2$ are inherited by the product nuclei, resulting in symmetric product nuclear spin states. Due to the exchange symmetry of the identical bosonic (fermionic) nuclei of K$_2$ (Rb$_2$), the rotational states are restricted to even (odd) parity.
\textbf{b}, The REMPI scheme used to identify rovibrational states of products. K$_2$ and Rb$_2$ molecules are state-selectively photoionized from the initial quantum state, $X^1\Sigma_g^+(v=0,N)$, via an intermediate state, $B^1\Pi_u(v',N')$. Here, $v$ and $v'$ are vibrational quantum numbers; $N$ and $N'$ are rotational quantum numbers. 
\textbf{c}, Timing diagram for product ionization and detection.
The optical trap confining the reactant KRb molecules is square-wave modulated at 7 kHz. During the dark phase of the modulation, K$_2$ and Rb$_2$ products are generated by the reaction, ionized using REMPI, and detected by ion mass spectrometry. The sequence is repeated for a duration of 1 s, until all reactants in the trap are depleted.
}
\label{fig1}
\end{figure}

In this study, we first show that nuclear spins are conserved throughout the ultracold reaction $^{40}\text{K}^{87}\text{Rb}$+$^{40}\text{K}^{87}\text{Rb}\rightarrow  \text{K}_2\text{Rb}_2^*\rightarrow $ $^{40}\text{K}_2$+$^{87}\text{Rb}_2$ (Fig.~\ref{fig1}a), where the K and Rb nuclei have high nuclear spins of $4$ and $3/2$, respectively, and the intermediate complex $\text{K}_2\text{Rb}_2^*$ has a long lifetime of 360(30) ns \cite{liu2020photo}. Because the products are homonuclear diatomic molecules, their rotational states are correlated with their nuclear spin states by the exchange symmetry of identical particles \cite{atkins2011molecular}. Using resonance-enhanced multiphoton ionization (REMPI) spectroscopy to probe the rotational states of the products, we find K$_2$ molecules predominantly in even-parity rotational states ($N_{\text{K}_2}=0,2,\cdots,12$) and Rb$_2$ in odd-parity states ($N_{\text{Rb}_2}=1,3,\cdots,19$), consistent with nuclear spin conservation. Utilizing this conservation, we demonstrate a technique to continuously alter the state distributions of the products by changing the superposition state of the reactant nuclear spins with an applied magnetic field. We model this behavior as a projection of the spin state of the reactant nuclei onto the symmetric or antisymmetric spin states of the product nuclei.

Our experiment begins with a trapped gas of $X^1\Sigma^+(v = 0,N_\text{KRb} = 0)$ ground-state KRb molecules prepared in a single hyperfine state.
Here $v$ and $N$ are vibrational and rotational quantum numbers, respectively. In each experimental cycle, we typically prepare $10^4$ molecules in a crossed optical dipole trap with a temperature of 500 nK and a peak density of $10^{12}$ cm$^{-3}$. Details of the apparatus regarding the production and detection of the gas are reported elsewhere~\cite{liu2020probing}.
In the presence of a sufficiently large magnetic bias field ($B \gtrsim 20$ G), the spins of the K and Rb nuclei are decoupled from one another, and the hyperfine state of the molecules can be written as
$|i_\text{K}=4,i_\text{Rb}=3/2,m_i^\text{K}=-4,m_i^\text{Rb}=1/2\rangle$, where $i$ and $m_i$ are the nuclear spins and their projections along the bias field, respectively~\cite{ospelkaus2010controlling, aldegunde2008hyperfine}. Once prepared, the molecules undergo the exchange reaction,
\begin{equation} \label{KRb reaction}
    2\text{KRb} (v = 0, N_\text{KRb} = 0) \rightarrow\ ^{40}\text{K}_2 (v = 0, N_{\textrm{K}_2}) +\,  ^{87}\text{Rb}_2 (v = 0, N_{\textrm{Rb}_2}),
\end{equation}
for which the product species were previously identified by combining single-photon ionization with ion mass spectrometry \cite{hu2019direct}. In order to investigate the behavior of nuclear spins throughout this reaction, we use state-controlled reactants at $B = 30$ G.

While directly probing the nuclear spin states of products is challenging due to their small energy splittings, we can infer this information from the parity of their rotational states.
For homonuclear diatomic molecules such as K$_2$ and Rb$_2$, nuclear spin and rotation are linked by the exchange symmetry of identical particles~\cite{atkins2011molecular}. 
In $^{40}$K$_2$, where the nuclei are bosonic, spin states that are symmetric under the exchange of nuclei require rotational states with even parity, while those that are antisymmetric require odd parity.
In $^{87}$Rb$_2$, where the nuclei are fermionic, this symmetry-parity correspondence is switched.
Therefore, if nuclear spins are conserved over the reaction, K$_2$ and Rb$_2$ will have symmetric spin states $|i_\text{K}=4,i_\text{K}=4,m_i^\text{K}=-4,m_i^\text{K}=-4\rangle$ and $|i_\text{Rb}=3/2,i_\text{Rb}=3/2,m_i^\text{Rb}=1/2,m_i^\text{Rb}=1/2\rangle$, respectively, allowing only even  rotational states of K$_2$ and odd rotational states of Rb$_2$ to be occupied.
This outcome is easily distinguishable from that where the nuclear spin states are statistically occupied. In this case, the population ratio between even and odd rotational states would be 5/4 for K$_2$ and 3/5 for Rb$_2$, as determined by the numbers of symmetric and antisymmetric spin states in each molecule \cite{atkins2011molecular}.

To detect the product rotational states, we use a $1+1'$ REMPI technique, which consists of an initial single-photon bound-to-bound transition from the electronic and vibrational ground-state $X^1\Sigma^{+}_g(v=0,N)$ to an electronically excited intermediate-state $B^1\Pi_u(v',N')$, followed by a single-photon bound-to-continuum transition that ionizes the molecules (Fig.~\ref{fig1}b). We drive the bound-to-bound transition using a frequency-tunable laser operating around 648 nm for the detection of K$_2$ and 674 nm for Rb$_2$. The bound-to-continuum transition is excited by a 532 nm pulsed laser for both product species. Figure \ref{fig1}c shows the timing diagram for product ionization and detection. Because the trap light can alter the outcome of the reaction~\cite{liu2020steering}, we apply a 7 kHz square wave modulation to its intensity, and probe the products during the dark phases of the modulation using REMPI pulses and ion mass spectrometry. In each cycle, ion signals are recorded until all the KRb reactants in the sample are depleted ($\sim$ 1 s). To avoid photoexcitation of the KRb molecules by the REMPI lasers~\cite{aikawa2010coherent,banerjee2012direct}, we shape the beams to have dark spots centered on the trapped sample (Extended Data Fig.~1).
Products, on the other hand, escape the trap because of their significant translational energies, and about $30\%$ of them are illuminated by the REMPI beams before they leave the detection region.

\begin{figure}
\centering
\includegraphics[width=5 in]{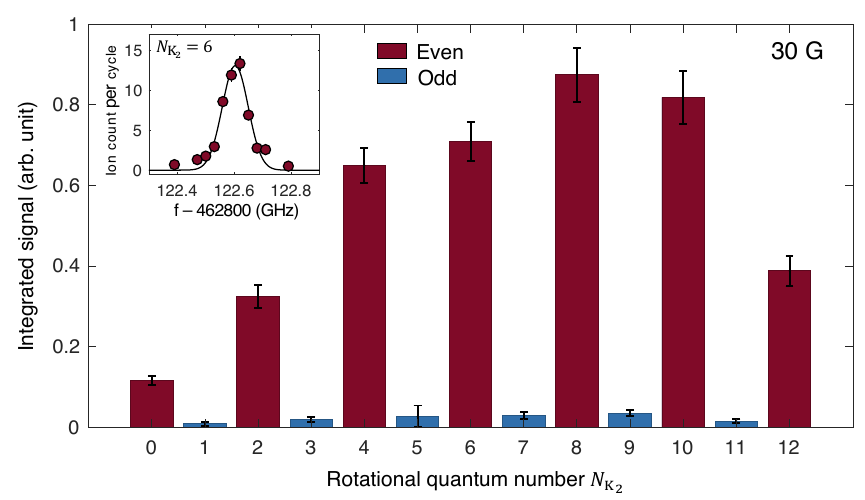}
\caption{\textbf{Rotational state occupation of $^{40}$K$_2$ products in a 30 G magnetic field.}
By scanning the 648 nm laser frequency to search for rotational lines within the $X^1\Sigma_g^+(v=0,N_{\text{K}_2})\rightarrow B^1\Pi_u(v'=1,N'_{\text{K}_2})$ vibronic band, we observe strong K$_2^+$ signals corresponding to rotational states with even values of $N_{\text{K}_2}$, and highly suppressed signals for odd values. The integrated signal for a given rotational state is obtained from a Gaussian function (inset, black curve) fitted to the measured REMPI resonance data (inset, red filled circles). Error bars denote the standard deviation of the mean (standard error). (Inset) A resonance signal obtained through REMPI spectroscopy for the transition $N_{\text{K}_{2}} = 6 \rightarrow N'_{\text{K}_{2}} = 6$; the Gaussian width is $\sigma=46\pm4$ MHz, and the error bars denote shot noise. The full REMPI spectrum is provided in Extended Data Fig.~2.
}
\label{fig2}
\end{figure}

\begin{figure}
\centering
\includegraphics[width=6 in]{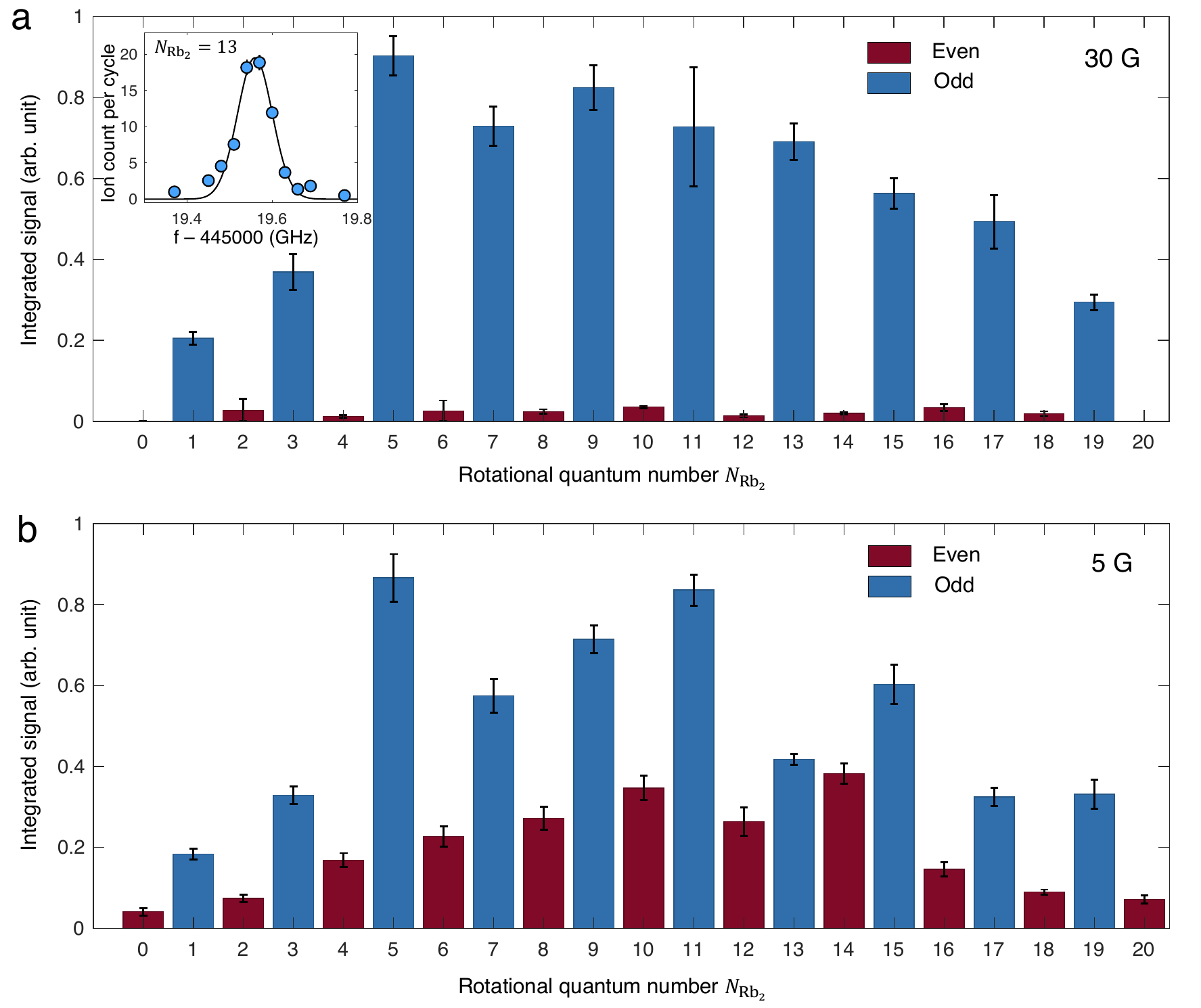}
    \caption{\textbf{Effect of an external magnetic field on the rotational state occupation of $^{87}$Rb$_2$ products.} \textbf{a}, Detected state occupation for a magnetic field of 30~G. By scanning the 674 nm laser frequency within the $X^1\Sigma_g^+(v=0,N_{\text{Rb}_2})\rightarrow B^1\Pi_u(v'=4,N'_{\text{Rb}_2})$ vibronic band, we observe strong Rb$_2^+$ signals for transitions from odd rotational states, and highly suppressed signals from even ones. The integrated signal for a given rotational state is obtained from a Gaussian function (inset, black curve) fitted to the measured REMPI resonance data (inset, blue filled circles). Error bars denote the standard deviation of the mean (standard error). (Inset) A resonance signal obtained through REMPI spectroscopy for the transition $N_{\text{Rb}_{2}} = 13 \rightarrow N'_{\text{Rb}_{2}} = 13$; the Gaussian width is $\sigma=43\pm3$ MHz, and the error bars denote shot noise. \textbf{b}, Detected state occupation for a magnetic field of 5~G. By scanning the laser frequency within the $X^1\Sigma_g^+(v=0,N_{\text{Rb}_2})\rightarrow B^1\Pi_u(v'=6,N'_{\text{Rb}_2})$ vibronic band, we observe strong Rb$_2^+$ signals for transitions from both even and odd rotational states. The full REMPI spectra at 30 and 5~G are given in Extended Data Figs.~3 and 4, respectively.
}
\label{fig3}
\end{figure}

To identify the occupied rotational states of the products, we scan the $648/674$ nm laser frequency to search for resonances around the $\Delta N=N'-N=0$ ($Q$ branch) transitions for $N>0$, and around the $\Delta N=1$ ($R$ branch) transition for $N=0$, and compare our measurements with theoretical calculations based on molecular potentials fitted by prior spectroscopic data.
We observe resonant signals corresponding to the states $N = 0-12$ for K$_2$ and $0 - 19$ for Rb$_2$ at frequencies that match the predicted values to within our measurement precision. The spectral widths of these resonances (Fig.~\ref{fig2} and Fig.~\ref{fig3}a insets) arise from the natural linewidth, the laser frequency uncertainty, and the Doppler width (Methods). We also search for states beyond $N_{\textrm{K}_2} = 12$ and $N_{\textrm{Rb}_2} = 19$, which are energetically forbidden based on the calculated reaction exothermicity of 9.54 cm$^{-1}$~\cite{yang2020global}, and find their populations to be consistent with zero.

In order to compare the signals from different rotational states for each product species, we fit each of the measured resonances with a Gaussian function, from which we obtain the total integrated signal of a given rotational state. The resulting signal distributions
are shown in Fig.~\ref{fig2} for K$_2$ and Fig.~\ref{fig3}a for Rb$_2$ as functions of the identified rotational quantum number, $N$.
Among those states that are energetically accessible, we find populations predominantly in the even-numbered rotational states of K$_2$ and odd-numbered rotational states of Rb$_2$.
This significantly deviates from statistical behavior, and provides direct evidence for the conservation of nuclear spins in a reaction involving high-spin nuclei and a long-lived intermediate complex. We note here, however, that the relative signal heights in Figs.~\ref{fig2} and \ref{fig3} are not yet sufficient to directly reconstruct the intrinsic product state distribution governed by the reaction dynamics, as there are uncertainties in the ionization efficiency associated with each state which arise from the REMPI beam geometry and the repetition rate for the detection (Methods).

Given the strong rotational parity preference observed at $B=30$ G, we can examine the effect of the magnetic field on the detected REMPI product signal by performing a similar measurement at $B=5$~G. 
To this end, we adiabatically ramp the field from 30 G to 5 G after the initial preparation of the reactants in the $| 4, 3/2, -4, 1/2\rangle$ state. 
We probe the state occupation of the Rb$_2$ products, and the resulting integrated signal distribution is shown in Fig.~\ref{fig3}b. Clear deviations from the $30$ G distribution are observed, as evidenced by the establishment of a significant population in even-numbered rotational states of Rb$_2$ at $5$ G. While such behavior naively seems to contradict the previous conclusion that nuclear spins are conserved throughout this ultracold reaction, this is not actually the case.
In fact, as we discuss below, these results are manifestations of nuclear spin conservation for a reactant hyperfine state which consists of a coherent superposition of several Zeeman sublevels. The ability to tune this superposition via an external magnetic field, along with the fact that nuclear spins remain unchanged throughout the reaction, are the keys to our control over the reaction outcome at the quantum state level.

\begin{figure}
\centering
\includegraphics[width=6.5 in]{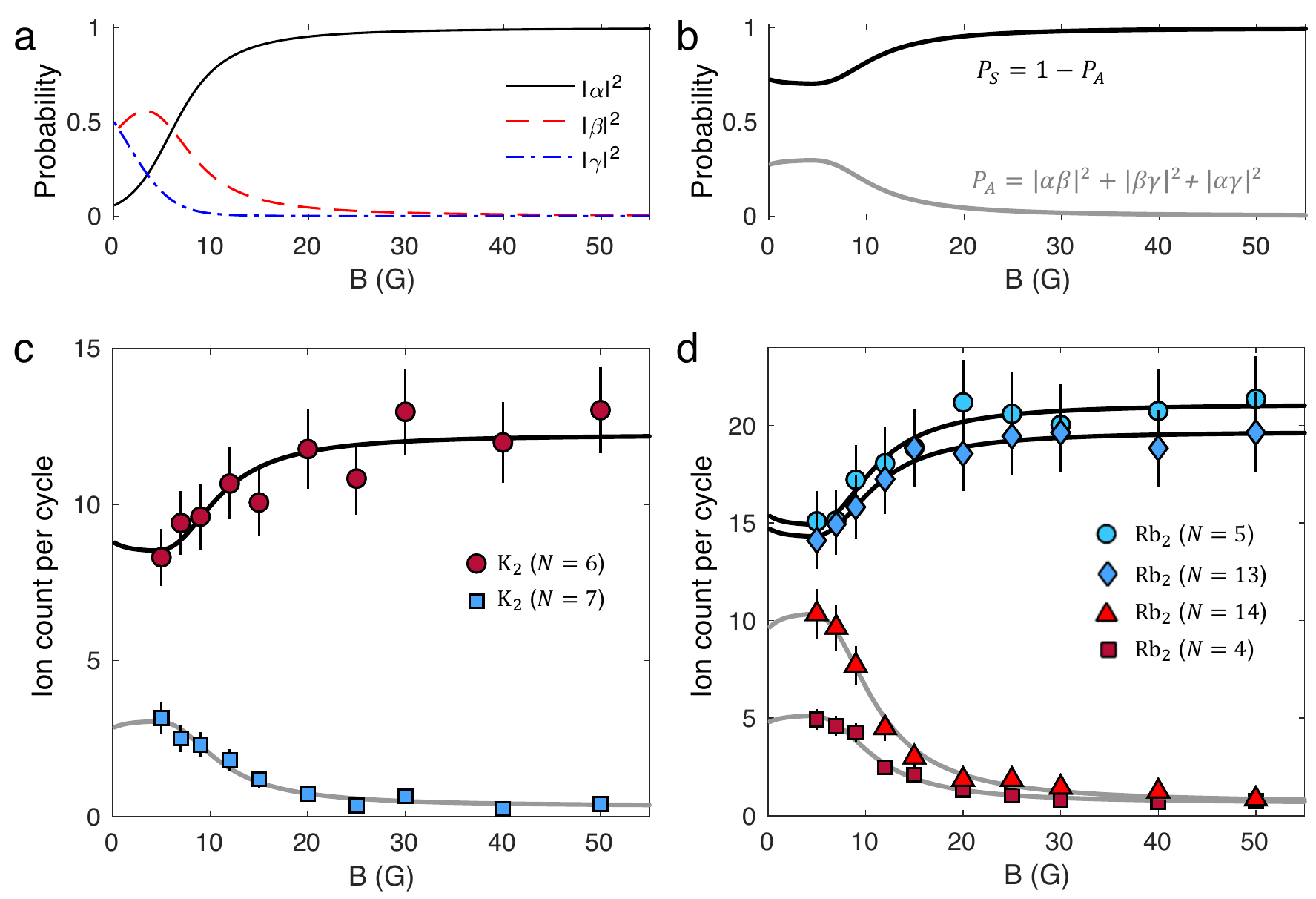}
        \caption{\textbf{Continuous control of product rotational state occupations with a magnetic field.} 
        \textbf{a}, Calculated admixture probabilities $|\alpha|^2$, $|\beta|^2$, and $|\gamma|^2$ for the reactant spin superposition (Eq. \ref{eq-psiKRb}) as a function of magnetic field. \textbf{b}, Probabilities for scattering into the symmetric ($P_{S}$) and antisymmetric ($P_{A}$) nuclear spin states of K$_2$ or Rb$_2$ based on the conservation of nuclear spins. 
        \textbf{c-d}, Product ion counts associated with the rotational states $N=6,7$ for K$_2$ (\textbf{c}) and $N=4,5,13,14$ for Rb$_2$ (\textbf{d}) obtained at different magnetic field values between 5 and 50 G. For each data set, the 648/674 nm laser frequency is fixed to be resonant with the corresponding transition, as identified from the spectra in Extended Data Figs.~2 and 3. The ion count for each data point is normalized by the corresponding number of experimental cycles ($\sim20$). Error bars include shot noise and $10\%$ ion number fluctuation due to the REMPI laser frequency uncertainty ($\sim 25$ MHz). Black (grey) curves are fits to the function $a\cdot P_{S(A)}+b$, where $a$ and $b$ are fit parameters representing the state-dependent proportionality constant and the offset arising from the detection noise floor, respectively. The typical root-mean-square error (RMSE) for the fit is 0.35.}
\label{fig4}
\end{figure}

We further explore this new form of control to continuously alter the distribution of quantum states of emerging product molecules by tuning the strength of the magnetic field in which the reaction occurs. As the magnitude of this field changes, so too does the nuclear spin superposition state of the KRb reactants, due to a competition between the Zeeman energy and the nuclear spin-spin interaction in the KRb hyperfine Hamiltonian~\cite{aldegunde2008hyperfine}. To realize this experimentally, we again initialize the molecules in the state $| 4, 3/2, -4, 1/2\rangle$, and subsequently ramp the field adiabatically to a final value $B$, where the nuclear spin state of the reactants takes the form,
\begin{equation}
    \psi_\text{KRb}=\alpha\left|4,\frac{3}{2},-4,\frac{1}{2}\right\rangle+\beta\left|4,\frac{3}{2},-3,-\frac{1}{2}\right\rangle+\gamma \left|4,\frac{3}{2},-2,-\frac{3}{2}\right\rangle.
    \label{eq-psiKRb}
\end{equation}
The admixture probabilities, $|\alpha|^2$, $|\beta|^2$, and $|\gamma|^2$, are calculated by diagonalizing the hyperfine Hamiltonian of KRb as a function of $B$ (Methods), as shown in Fig.~\ref{fig4}a. For $B \gtrsim 20$ G, one has $|\alpha|^2 \approx 1 $, and the reactant state is dominated by $|4,3/2,-4,1/2\rangle$.  For $B \lesssim 20$ G,  the state becomes significantly mixed with the other two spin components. With this form of the input state, we probe the outcome of the reaction for values of $B$ in the range $5-50$ G. Without loss of generality, we measure changes in the amplitudes of the resonances corresponding to the rotational states $N = 6,7$ for K$_2$ and $N = 4,5,13,14$ for Rb$_2$ (Fig.~\ref{fig4}c,d). We observe that the odd (even) states of K$_2$ (Rb$_2$), which are unoccupied at higher magnetic fields, acquire significant population as $B$ decreases, while the population in the even (odd) states is suppressed. This shows that the relative occupation of odd and even parity states of the products can be altered continuously by controlling the nuclear spin state of the reactants via an external field.

This behavior can be explained using a model based on angular momentum algebra, which we derive in accordance with the conservation of nuclear spins throughout the reaction (Methods). Given the form of the reactant input state (Eq.~\ref{eq-psiKRb}), the probabilities for scattering into the antisymmetric ($A$) and symmetric ($S$) spin states of K$_2$ or Rb$_2$ are $P_A=|\alpha \beta|^2+|\alpha \gamma|^2+|\beta \gamma|^2$ and $P_S=1-P_A$, respectively. 
The magnetic field dependence of these probabilities, shown in Fig.~\ref{fig4}b, derives from that of the admixture coefficients (Fig.~\ref{fig4}a).
Because of the exchange symmetry of identical nuclei, the detected populations of even (odd) rotational states are proportional to $P_S$ ($P_A$) for K$_2$ and $P_A$ ($P_S$) for Rb$_2$. The constant of proportionality depends on the rotational state distribution resulting from the reaction dynamics, as well as the state-dependent ionization efficiency, and is  unique for each rotational state.
As a result, we fit each data set in Fig. \ref{fig4}c,d to the function $a\cdot P_{S/A} + b$, with free parameters $a$ and $b$. Here, $a$ represents the aforementioned proportionality constant,
and $b$ is an offset arising from the noise floor of the detection.
The fitted curves (Fig.~\ref{fig4}c,d) show that our model for $P_{S/A}$ describes well the magnetic field dependence of the rotational state populations.

While this model was derived for a specific input state, Eq.~(\ref{eq-psiKRb}), the methodology does not depend on the particular form of this state, and can be generalized to account for arbitrary states of the reactant nuclear spins. Experimentally, such states may be realized using a combination of external magnetic fields and microwave control techniques~\cite{ospelkaus2010controlling,park2017second}. We also note that the sensitivity of the rotational state occupation to external fields observed here is not a feature that is unique to the KRb $+$ KRb system, but can also play a significant role in other chemical reactions involving identical nuclei, such as those found in atmospheric~\cite{gunthardt2019evidence} and interstellar processes~\cite{lique2014ortho}. Our results therefore indicate that the effects of ambient fields may warrant consideration in future studies of reactions in these environments.


In summary, we have demonstrated control over the product state occupation of a chemical reaction by manipulating conserved nuclear spins via an external magnetic field. We show that the complexity of the dynamics for complex-forming reactions does not fundamentally limit their controllability at the quantum state-to-state level. An important advantage of this technique is that the quantum coherence between different spin-components of the reactants is inherited by the products. Such an effect can be utilized to engineer correlations between the states of product molecules originating from the same reaction, which can enable the generation of quantum entanglement between reaction products~\cite{gong2003entanglement,li2019entanglement}. Moreover, coherent control over reactions between identical diatomic molecules is particularly well-suited for the detection of quantum interference between different reaction pathways~\cite{brumer2000identical}.

\clearpage

\bibliography{refs}
\bibliographystyle{naturemag}
\newpage
\setcounter{figure}{0}    

\begin{center}
\textbf{Methods}
\end{center}
\textbf{1+1\ensuremath{'} REMPI spectroscopy}
Our REMPI scheme for the state-selective detection of K$_2$ (Rb$_2$) utilizes two lasers: a continuous-wave external cavity diode laser operating at 648 (674) nm, and a pulsed Nd:YVO$_4$ laser with a frequency-doubled output wavelength of 532 nm and a pulse duration of $\sim$ 10 ns.
During an experiment, temporally-overlapping light pulses derived from these lasers are repeatedly applied to ionize reaction products. The duration of the 648 (674) nm pulse is controlled by an acoustic-optical modulator (AOM) to be $\sim 40$ ns, and the relative delay between the two pulses is scanned to maximize the resulting ion signals. The optical power in the 648 and 674 nm beams is 8 and 15 mW, respectively, sufficient to saturate the targeted transitions.
The pulse energy (optical power) of the 532 nm laser is set to be 50 $\mu$J (350 mW), resulting in the ionization of $> 80\%$ of the electronically-excited molecules. 
The frequencies of the diode lasers are stablized to an accuracy of $25$~MHz through locking to a laser wavelength meter (Bristol 771A).

To avoid excitation of the reactant KRb molecules by the REMPI lasers, we generate dark spots in the centers of the beams using the optical setup illustrated in Extended Data Fig.~1. The 648/674 nm and 532 nm beams are first combined on a dichroic mirror, and are then sent through a dark mask (Thorlabs R1DF100). The mask is located in the object plane of an achromatic lens ($f = 300$~mm), and is projected onto the image plane coinciding with the molecular cloud location, resulting in the beam profiles shown in the insets. The dark spot size is chosen based on the KRb cloud size, which is up to 80 $\mu$m ($4\sigma$ width) in its longest direction. The length of the dark region along the direction of propagation is approximately 7 mm on either side of the image plane (defined as when the optical intensity inside the dark spot increases to half of the maximal value measured in the image plane).

For $^{40}$K$_2$, the bound-to-bound transition lines we target are from the $X^1\Sigma_g^+(v=0,N)\rightarrow B^1\Pi_u(v'=1,N')$ vibronic band.
The transition dipole moment for these lines is calculated to be $(1.6\ e\cdot a_0)\sqrt{S_N^{P,Q,R}}$. Here, $e$ is the electron charge, $a_0$ is the Bohr radius, and $S_N^{P,Q,R}$ is the H\"{o}nl-London factor which takes different forms for $P$, $Q$, and $R$ branch rotational transitions.
For $\Sigma\rightarrow\Pi$ electronic transitions, they are given by $S_N^P=\frac{N-1}{2N+1}$, $S_N^Q=1$, and $S_N^R=\frac{N+2}{2N+1}$~\cite{bernath2016spectra}.
We calculate the linewidth (lifetime) of the excited state to be 14.5 MHz (11 ns), close to the measured value of 13 MHz (12.2 ns) reported in Ref.~\cite{lemont1977radiative}.
An additional contribution to the experimentally measured linewidth is Doppler broadening arising from the large translational energy of K$_2$ products. For K$_2$ molecules moving at the maximum velocity allowed by the reaction exothermicity, the Doppler contribution to the linewidth is $\sigma_D\sim40$ MHz.
The REMPI spectroscopy in this work requires knowing transition frequencies better than $0.001$~cm$^{-1}$ (or 30~MHz), which is not available from the literature.
To identify them, our initial search of the transition lines is guided by frequencies that are calculated using molecular potentials from Ref.~\cite{falke2006sigma,engelke1984k2}.

For $^{87}$Rb$_2$, the bound-to-bound transition lines we used are from the $X^1\Sigma_g^+(v=0,N)\rightarrow B^1\Pi_u(v'=4,N')$ and $X^1\Sigma_g^+(v=0,N)\rightarrow B^1\Pi_u(v'=6,N')$ vibronic bands, which have transition dipole moments that are calculated to be $(1.5\ e\cdot a_0)\sqrt{S_N^{P,Q,R}}$ and $(0.9\ e\cdot a_0)\sqrt{S_N^{P,Q,R}}$, respectively. The natural linewidth (lifetime) of the excited state is calculated to be $13.9$~MHz (11~ns).
For Rb$_2$ molecules moving at the maximum allowed velocity, the Doppler contribution to the linewidth is $\sigma_D\sim30$ MHz.
Our initial search of the transition lines is guided by frequencies that are calculated using molecular potentials from Ref. ~\cite{seto2000direct,caldwell1980high,amiot1997optical}. 

For the product photoionization efficiency, one common factor to consider is the state-dependent rate of the bound-to-bound transition used for REMPI.
Here, we circumvent this by exploiting the uniformity of the $Q$ branch transitions for almost all available rotational states, and utilizing sufficient laser power to saturate each transition. For our experimental setup, an additional factor that affects the ionization efficiency comes from the unique product velocity distribution associated with each rotational state. This distribution arises from the partitioning of the exothermic energy into the rotations and translations of K$_2$ and Rb$_2$ by the underlying dynamics of the reaction, and is not known \textit{a priori}. Due to the finite beam size and limited repetition rate of the REMPI lasers (Extended Data Fig.~1), lower velocity products are more likely to be ionized before escaping the detection region compared to those with higher velocities. Although this effect is difficult to quantify in the current iteration of the experiment, it can be eliminated in the future by increasing the repetition rate of the REMPI lasers and/or enlarging the detection region, which will allow all products to be illuminated before they escape.

\textbf{State-decomposition calculation for reactant molecules}
To obtain the coefficients $\alpha$, $\beta$, and $\gamma$ of the state-decomposition given in Eq.~(\ref{eq-psiKRb}) of the main text, we diagonalize the molecular Hamiltonian for the $^{40}$K$^{87}$Rb reactants using the basis of uncoupled hyperfine states. If we include the rotational degrees of freedom, we can label these basis states by $\left|N,m_{N},i_{\text{K}},m_{i}^{\text{K}},i_{\text{Rb}},m_{i}^{\text{Rb}}\right>$, where $N$ represents the rotational angular momentum quantum number, $m_{N}$ is its projection onto the $z$-axis, $i_{\text{K,Rb}}$ are the nuclear spins of the K and Rb atoms, and $m_{i}^{\text{K,Rb}}$ are the corresponding projections onto the $z$-axis. The direction of the $z$-axis here is defined by an externally applied magnetic field. In the electronic ground state, $X^{1}\Sigma^{+}$, as well as the vibrational ground state, the molecular Hamiltonian in the presence of external electric and magnetic fields can be expressed as~\cite{Brown2003,aldegunde2008hyperfine},
\begin{equation}
    H=H_{\text{rot}}+H_{\text{HF}}+H_{\text{S}}+H_{\text{Z}}.
\label{MolecularHamiltonian}
\end{equation}
Here,
\begin{equation}
    H_{\text{rot}}=B_{\text{rot}}\textit{\textbf{N}}^{2},
\label{RotationalEnergy}
\end{equation}
\begin{equation}
    H_{\text{HF}}=-e\left(\boldsymbol{\nabla}\textbf{E}\right)_{\text{K}}\cdot\textbf{Q}_{\text{K}}-e\left(\boldsymbol{\nabla}\textbf{E}\right)_{\text{Rb}}\cdot\textbf{Q}_{\text{Rb}}+c_{\text{K}}\textit{\textbf{N}}\cdot\textit{\textbf{I}}_{\text{K}}+c_{\text{Rb}}\textit{\textbf{N}}\cdot\textit{\textbf{I}}_{\text{Rb}}+c_{4}\textit{\textbf{I}}_{\text{K}}\cdot\textit{\textbf{I}}_{\text{Rb}},
\label{HyperfineEnergy}
\end{equation}
\begin{equation}
    H_{\text{S}}=-\boldsymbol{\mu}\cdot\textit{\textbf{E}},
\label{StarkEnergy}
\end{equation}
\begin{equation}
    H_{\text{Z}}=-g_{r}\mu_{N}\textit{\textbf{N}}\cdot\textit{\textbf{B}}-g_{\text{K}}\mu_{N}\left(1-\sigma_{\text{K}}\right)\textit{\textbf{I}}_{\text{K}}\cdot\textit{\textbf{B}}-g_{\text{Rb}}\mu_{N}\left(1-\sigma_{\text{Rb}}\right)\textit{\textbf{I}}_{\text{Rb}}\cdot\textit{\textbf{B}},
\label{ZeemanEnergy}
\end{equation}
where $\textit{\textbf{N}}$ is the rotational angular momentum operator, $\textit{\textbf{I}}_{\text{K,Rb}}$ are the nuclear spin operators for the K and Rb nuclei, respectively, $\left(\boldsymbol{\nabla}\textbf{E}\right)_{\text{K,Rb}}$ are the intramolecular electric field gradients at the K and Rb nuclei, respectively, $e\textbf{Q}_{\text{K,Rb}}$ are the nuclear electric quadrupole moment operators for K and Rb, respectively, $\boldsymbol{\mu}$ is the molecular dipole moment, $\textit{\textbf{E}}$ is the external electric field, and $\textit{\textbf{B}}$ is the external magnetic field.

In this molecular Hamiltonian, Eq.~(\ref{RotationalEnergy}) denotes the rotational contribution to the energy, with corresponding rotational constant $B_{\text{rot}}/h=1.1139514\,\text{GHz}$~\cite{ospelkaus2010controlling,Neyenhuis2012}. Eq.~(\ref{HyperfineEnergy}) represents the hyperfine energy, where $-e\left(\boldsymbol{\nabla}\textbf{E}\right)_{i}\cdot\textbf{Q}_{i}$ describes the interaction between the intramolecular electric field gradient at nucleus $i$ and the corresponding nuclear electric quadrupole moment, which is characterized by the electric quadrupole coupling constants $\left(eqQ\right)_{\text{K}}/h=0.452\,\text{MHz}$ and $\left(eqQ\right)_{\text{Rb}}/h=-1.308\,\text{MHz}$~\cite{Neyenhuis2012}. The remaining three terms in Eq.~(\ref{HyperfineEnergy}) describe the interactions of the individual nuclear spins with the magnetic field associated with the rotation of the molecule, where the corresponding coupling constants are given by $c_{\text{K}}/h=-24.1\,\text{Hz}$ and $c_{\text{Rb}}/h=420.1\,\text{Hz}$~\cite{aldegunde2008hyperfine}, as well as the scalar nuclear spin-spin interaction, with an associated coupling constant $c_{4}/h=-2030.4\,\text{Hz}$~\cite{aldegunde2008hyperfine}. The Stark Hamiltonian given by Eq.~(\ref{StarkEnergy}) describes the interaction between the permanent electric dipole moment of KRb, $\mu=0.574\,\text{Debye}$~\cite{NiThesis2009}, and an external electric field. The Zeeman contribution to the energy (Eq.~(\ref{ZeemanEnergy})), on the other hand, consists of three separate terms: the interaction between the external magnetic field and the magnetic moment arising from molecular rotation, with a corresponding rotational $g$-factor of $g_{r}=0.014$~\cite{aldegunde2008hyperfine}, and the interactions between the magnetic moments of the individual nuclei, whose $g$-factors are given by $g_{\text{K}}=-0.324$ and $g_{\text{Rb}}=1.834$~\cite{aldegunde2008hyperfine}, and the external magnetic field. The remaining parameters in Eq.~(\ref{ZeemanEnergy}), $\mu_{N}$ and $\sigma_{\text{K,Rb}}$, represent the nuclear magneton and the nuclear shielding constants, respectively. For the K and Rb nuclei, the nuclear shielding constants are $\sigma_{\text{K}}=1321\,\text{ppm}$ and $\sigma_{\text{Rb}}=3469\,\text{ppm}$~\cite{aldegunde2008hyperfine}.

For the reactant state-decomposition calculations shown in the main text (Fig.~\ref{fig4}a), we include rotational states up to $N_{\text{max}}=1$ when diagonalizing this Hamiltonian in the basis of uncoupled hyperfine states, $\left|N,m_{N},i_{\text{K}},m_{i}^{\text{K}},i_{\text{Rb}},m_{i}^{\text{Rb}}\right>$, where $i_{\text{K}}=4$ and $i_{\text{Rb}}=3/2$. We also make use of the fact that the ground-state molecules are initially prepared at zero electric field and at a high magnetic field, $\textit{B}\approx544\,\text{G}$, where the uncoupled basis states are a good representation of the eigenstates of the molecular Hamiltonian (Eq.~(\ref{MolecularHamiltonian})). In this configuration, the KRb molecules are produced in their rotational, vibrational, and electronic ground state, and in the hyperfine state $\left|\psi_{\text{KRb}}^{i}\right>\equiv\left|N=0,m_{N}=0,i_{\text{K}}=4,m_{i}^{\text{K}}=-4,i_{\text{Rb}}=3/2,m_{i}^{\text{Rb}}=1/2\right>$~\cite{liu2020probing}. This corresponds to the situation where $\alpha\approx1$ and $\beta=\gamma\approx0$ in Eq.~(\ref{eq-psiKRb}) of the main text. As the electric and magnetic fields are changed adiabatically following the initial preparation of the reactant molecules, we obtain the coefficients $\alpha$, $\beta$, and $\gamma$ by diagonalizing the molecular Hamiltonian (Eq.~(\ref{MolecularHamiltonian})) at the desired values of the electric and magnetic fields, and selecting the eigenstate which is adiabatically connected to the initial hyperfine state, $\left|\psi_{\text{KRb}}^{i}\right>$. The coefficients of the expansion of this eigenstate in terms of the uncoupled basis states are then $\alpha$, $\beta$, and $\gamma$. For the calculations shown in Fig.~\ref{fig4}a of the main text, we set $\textit{E}=0$ for simplicity, as the small electric field used in the experiment, $\textit{E}=18\,\text{V}/\text{cm}$, has a negligible effect on the state-decomposition, and we numerically calculate $\alpha$, $\beta$, and $\gamma$ as a function of the magnetic field.

\textbf{Selection rules for nuclear spin conservation in ultracold reactions}
Selection rules in chemical reactions arise when the $S$-matrix is diagonalizable in a basis made up of the eigenstates of some conserved quantity, also known as a collision constant~\cite{quack1977detailed}. Due to their weak couplings to other degrees of freedom, nuclear spins of molecules may be conserved in chemical reactions~\cite{oka2004nuclear}, and so their eigenstates may be used to derive one such selection rule. For ultracold reactions of bi-alkali molecules of the form, $\text{AB}+\text{AB}\rightarrow \text{A}_2+\text{B}_2$, the nuclear spin states of the heteronuclear reactants can be described in the uncoupled basis, $\{|i_\text{A1},i_\text{B1},m_\text{A1},m_\text{B1}\rangle\otimes50|i_\text{A2},i_\text{B2},m_\text{A2},m_\text{B2}\rangle\}$, while those of the homonuclear products are more naturally expressed in the coupled basis, $\{|I_{\text{A}_2},M_{\text{A}_2};i_\text{A1},i_\text{A2}\rangle\otimes |I_{\text{B}_2},M_{\text{B}_2};i_\text{B1},i_\text{B2}\rangle\}$, in which the states obey the exchange symmetry of identical nuclei. 
Here, A1 (A2) and B1 (B2) denote atomic nuclei from the 1st (2nd) AB molecules in the reaction. The total spin is defined as $\mathbf{I}_{\text{A}_2(\text{or }\text{B}_2)}=\mathbf{i}_\text{A1(or B1)}+\mathbf{i}_\text{A2(or B2)}$ and its projection is $M_{\text{A}_2(\text{or }\text{B}_2)}=m_\text{A1(or B1)}+m_\text{A2(or B2)}$. The coupled basis state $|I,M;i_1,i_2\rangle$ is symmetric under interchange of particles 1 and 2 if $i_1+i_2-I$ is even, and antisymmetric if $i_1+i_2-I$ is odd.

To obtain the probabilities for scattering into different channels, or product states, we can transform the reactant nuclear spin state from the uncoupled basis into the coupled basis. The coefficients of the reactant state in the coupled basis then provide the desired information regarding the branching into the different reaction channels. This basis transformation can be performed using the general formula for the addition of angular momentum, 
\begin{equation}\label{eq-CG}
    |i_1,i_2,m_1,m_2\rangle = \sum_{I=|i_1-i_2|}^{i_1+i_2}\langle I,M;i_1,i_2|i_1,i_2,m_1,m_2\rangle |I,M;i_1,i_2\rangle,
\end{equation}
where  $M=m_1+m_2$ and $\langle I,M;i_1,i_2|i_1,i_2,m_1,m_2\rangle$ are the Clebsch-Gordan (CG) coefficients.  
Take the reactant state
\begin{equation}    
    \psi_\text{KRb}=\alpha\left|4,\frac{3}{2},-4,\frac{1}{2}\right\rangle+\beta\left|4,\frac{3}{2},-3,-\frac{1}{2}\right\rangle+\gamma \left|4,\frac{3}{2},-2,-\frac{3}{2}\right\rangle,
    \label{eq-psiKRb-S}
\end{equation}
for example. Assuming that nuclear spins are spectators during the reaction, we can write the reactant nuclear spin state as,
\begin{eqnarray}
    &\psi_\text{KRb}&\otimes\psi_\text{KRb} = 
    \alpha^2\left|4,\frac{3}{2},-4,\frac{1}{2}\right\rangle\otimes \left|4,\frac{3}{2},-4,\frac{1}{2}\right\rangle+\beta^2\left|4,\frac{3}{2},-3,-\frac{1}{2}\right\rangle\otimes\left|4,\frac{3}{2},-3,-\frac{1}{2}\right\rangle \nonumber \\[10pt]
    &&+\gamma^2\left|4,\frac{3}{2},-2,-\frac{3}{2}\right\rangle\otimes\left|4,\frac{3}{2},-2,-\frac{3}{2}\right\rangle +\{\cdots\} \nonumber\\[10pt]
    &=&
    \alpha^2\left|4,4,-4,-4\right\rangle\otimes \left|\frac{3}{2},\frac{3}{2},\frac{1}{2},\frac{1}{2}\right\rangle+\beta^2\left|4,4,-3,-3\right\rangle\otimes\left|\frac{3}{2},\frac{3}{2},-\frac{1}{2},-\frac{1}{2}\right\rangle \nonumber \\[10pt]
    &&+\gamma^2\left|4,4,-2,-2\right\rangle\otimes\left|\frac{3}{2},\frac{3}{2},-\frac{3}{2},-\frac{3}{2}\right\rangle +\{\cdots\} \nonumber\\[10pt]
    &=&\alpha^2|8,-8\rangle\otimes\left(\sqrt{\frac{3}{5}}|3,1\rangle-\sqrt{\frac{2}{5}}|1,1\rangle\right) \label{eq-A0}\\[10pt]
    &&+\beta^2\left(-\sqrt{\frac{7}{15}}|6,-6\rangle+\sqrt{\frac{8}{15}}|8,-6\rangle\right)\otimes\left(\sqrt{\frac{3}{5}}|3,-1\rangle-\sqrt{\frac{2}{5}}|1,-1\rangle\right)\\[10pt]
    &&+\gamma^2\left(\sqrt{\frac{45}{143}}|4,-4\rangle-\sqrt{\frac{14}{55}}|6,-4\rangle+\sqrt{\frac{28}{65}}|8,-4\rangle\right)\otimes|3,-3\rangle	\\[10pt]
    &&+\alpha\beta|8,-7\rangle\otimes\left(\sqrt{\frac{9}{10}}|3,0\rangle-\sqrt{\frac{1}{10}}|1,0\rangle\right)		\\[10pt]
    &&+\alpha\gamma\left(\sqrt{\frac{8}{15}}|6,-6\rangle+\sqrt{\frac{7}{15}}|8,-6\rangle\right)\otimes\left(\sqrt{\frac{3}{5}}|1,-1\rangle+\sqrt{\frac{2}{5}}|3,-1\rangle\right)	\\[10pt]
    &&+\beta\gamma\left(\frac{2}{\sqrt{5}}|8,-5\rangle-\frac{1}{\sqrt{5}}|6,-5\rangle\right)\otimes|3,-2\rangle     \\[10pt]
    &&+\alpha\beta|7,-7\rangle\otimes\sqrt{\frac{1}{2}}\left(|0,0\rangle-|2,0\rangle\right) \label{eq-A1}\\[10pt]
    &&-\alpha\gamma|7,-6\rangle\otimes|2,-1\rangle  \label{eq-A2}\\[10pt]
    &&+\beta\gamma\left(\sqrt{\frac{9}{13}}|5,-5\rangle-\sqrt{\frac{4}{13}}|7,-5\rangle\right)\otimes|2,-2\rangle,   \label{eq-A3}
\end{eqnarray}
where we have used Eq. (\ref{eq-CG}) to obtain the final equality, and $\{\cdots\}$ represents all additional terms. The last step gives the reactant state in the coupled basis, where we omit values of $i_\text{K}$ and $i_\text{Rb}$ for convenience. Each of the resulting tensor product states in Eqs. (\ref{eq-A0}-\ref{eq-A3})  corresponds to a possible scattered channel. Among them, the terms in Eqs. (\ref{eq-A1}-\ref{eq-A3}) are antisymmetric under the exchange of identical nuclei, while all others are symmetric. Therefore, the total probability for scattering into an antisymmetric nuclear spin state of the products is given by $P_A=|\alpha \beta|^2+|\alpha \gamma|^2+|\beta \gamma|^2$, whereas that for scattering into a symmetric state is $P_S=1-P_A$. 
Furthermore, based on the exchange symmetry of identical nuclei, the populations of even (odd) rotational states of the products are proportional  to $P_S$ ($P_A$)  when the nuclei are bosonic, and are proportional to $P_A$ ($P_S$) when the nuclei are fermionic.

While this result was derived for a specific input state (Eq.~(\ref{eq-psiKRb-S})), the methodology does not depend on the particular form of this state, so that the formulae for $P_A$ and $P_S$ can be generalized to account for any initial states of the reactants. Additionally, as far as the summed probabilities are concerned, it is not strictly necessary  to  consider  the  CG  coefficients  of  both reactant and product dressed  states. Instead, it is sufficient, for the purposes of this work, to consider only the CG coefficients of the dressed reactant states.
However, for a state-to-state measurement which resolves individual nuclear spin states, it is, strictly speaking, necessary to consider the CG coefficients of the dressed states. This will be reported in another work.

\clearpage

\textbf{Data availability} The data that support the findings of this study are available from the corresponding author upon reasonable request.

\bigskip

\textbf{Code availability} The computer code used to analyse the data is available from the corresponding author upon reasonable request.

\bigskip

\textbf{Acknowledgements} We thank T. Rosenband for providing the code used to calculate the molecular hyperfine structure, J. Bohn for discussions,  R. Vexiau for calculations of relevant molecular transitions, and N. Hutzler and H. Guo for critical readings of the manuscript. This work is supported by  the DOE Young Investigator Program, the David and Lucile Packard Foundation, and the NSF through the Harvard-MIT CUA.  M.A.N. is supported by a HQI fellowship.  G.Q. acknowledges funding from the FEW2MANY-SHIELD Project No. ANR-17-CE30-0015 from Agence Nationale de la Recherche. 

\bigskip

\textbf{Author contributions} M.-G.H., Y.L., M.A.N., L.Z., and K.-K.N. carried out the experimental work and data analysis. M.-G.H., G.Q., and O.D. performed the theoretical work. All authors contributed to interpreting the results and writing the manuscript. 

\bigskip

\textbf{Competing interests} The authors declare that they have no competing financial interests. 

\bigskip

\textbf{Corresponding author} \\Correspondence to Ming-Guang Hu (e-mail: mingguanghu@g.harvard.edu)


\clearpage

\textbf{Extended Data Figure 1. The optical setup for generating the REMPI beams.} 
The 648/674 nm and 532 nm lasers are combined on a dichroic mirror, and are then sent through a dark mask (Thorlabs R1DF100) and an achromatic lens ($f=300$~mm). The resulting beam profiles in the image plane have a 1~mm outer diameter and a $100~\mu$m inner diameter.

\textbf{Extended Data Figure 2. REMPI spectrum for $^{40}$K$_2$ product molecules at 30 G.} By scanning the 648 nm laser frequency to search for rotational lines within the $X^1\Sigma_g^+(v=0,N_{\text{K}_2})\rightarrow B^1\Pi_u(v'=1,N'_{\text{K}_2})$ vibronic band, we observe strong K$_2^+$ signals at frequencies corresponding to rotational states with even values of $N_{\text{K}_2}$ (red filled circles), and highly suppressed signals for odd values (black open circles).
The ion count for each data point is normalized by the corresponding number of experimental cycles ($\sim16$); the error bars denote shot noise.
For $N_{\text{K}_2}>0$ we drive transitions with $N'_{\text{K}_{2}}-N_{\text{K}_{2}}=0$ ($Q$ branch), whereas for $N_{\text{K}_2}=0$, we drive the only allowed transition, with $N'_{\text{K}_{2}}-N_{\text{K}_{2}}=1$ ($R$ branch).
Blue dashed lines indicate the predicted transition frequencies. 
We do not observe any signals at frequencies corresponding to states with $N_{\text{K}_2} > 12$. 
Gaussian fits (black curves) are applied to each signal peak, yielding a typical spectral linewidth (1$\sigma$) of $\sim50$ MHz. 

\textbf{Extended Data Figure 3. REMPI spectrum for $^{87}$Rb$_2$ product molecules at 30 G.} The frequency of the 674 nm laser is scanned within the $X^1\Sigma_g^+(v=0,N_{\text{Rb}_2})\rightarrow B^1\Pi_u(v'=4,N'_{\text{Rb}_2})$ vibronic band. We observe strong Rb$_2^+$ signals for transitions from odd rotational states (blue filled circles), and highly suppressed signals from even ones (black open circles). The ion count for each data point is normalized by the corresponding number of experimental cycles ($\sim16$); the error bars denote shot noise. We drive $Q$ branch transitions for $N_{\text{Rb}_2}>0$, and $R$ branch for $N_{\text{Rb}_2} = 0$. Blue dashed lines indicate the predicted transition frequencies.  We do not observe any signals at frequencies corresponding to states with $N_{\text{Rb}_2} > 19$. Gaussian fits (black curves) are applied to each signal peak, yielding a typical spectral linewidth (1$\sigma$) of $\sim40$ MHz.

\textbf{Extended Data Figure 4. REMPI spectrum for $^{87}$Rb$_2$ product molecules at 5 G.} The frequency of the 674 nm laser is scanned within the $X^1\Sigma_g^+(v=0,N_{\text{Rb}_2})\rightarrow B^1\Pi_u(v'=6,N'_{\text{Rb}_2})$ vibronic band. We observe strong Rb$_2^+$ signals for transitions from both even (red filled circles) and odd (blue filled circles) rotational states. The ion count for each data point is normalized by the corresponding number of experimental cycles ($\sim20$); the error bars denote shot noise. We drive $Q$ branch transitions for $N_{\text{Rb}_2}>0$, and $R$ branch for $N_{\text{Rb}_2} = 0$. Blue dashed lines indicate the predicted transition frequencies.  We do not observe any signals at frequencies corresponding to states with $N_{\text{Rb}_2} > 19$. Gaussian fits (black curves) are applied to each signal peak, yielding a typical spectral linewidth (1$\sigma$) of $\sim40$ MHz. 

\clearpage

\begin{figure}
\centering
\includegraphics[width=5 in]{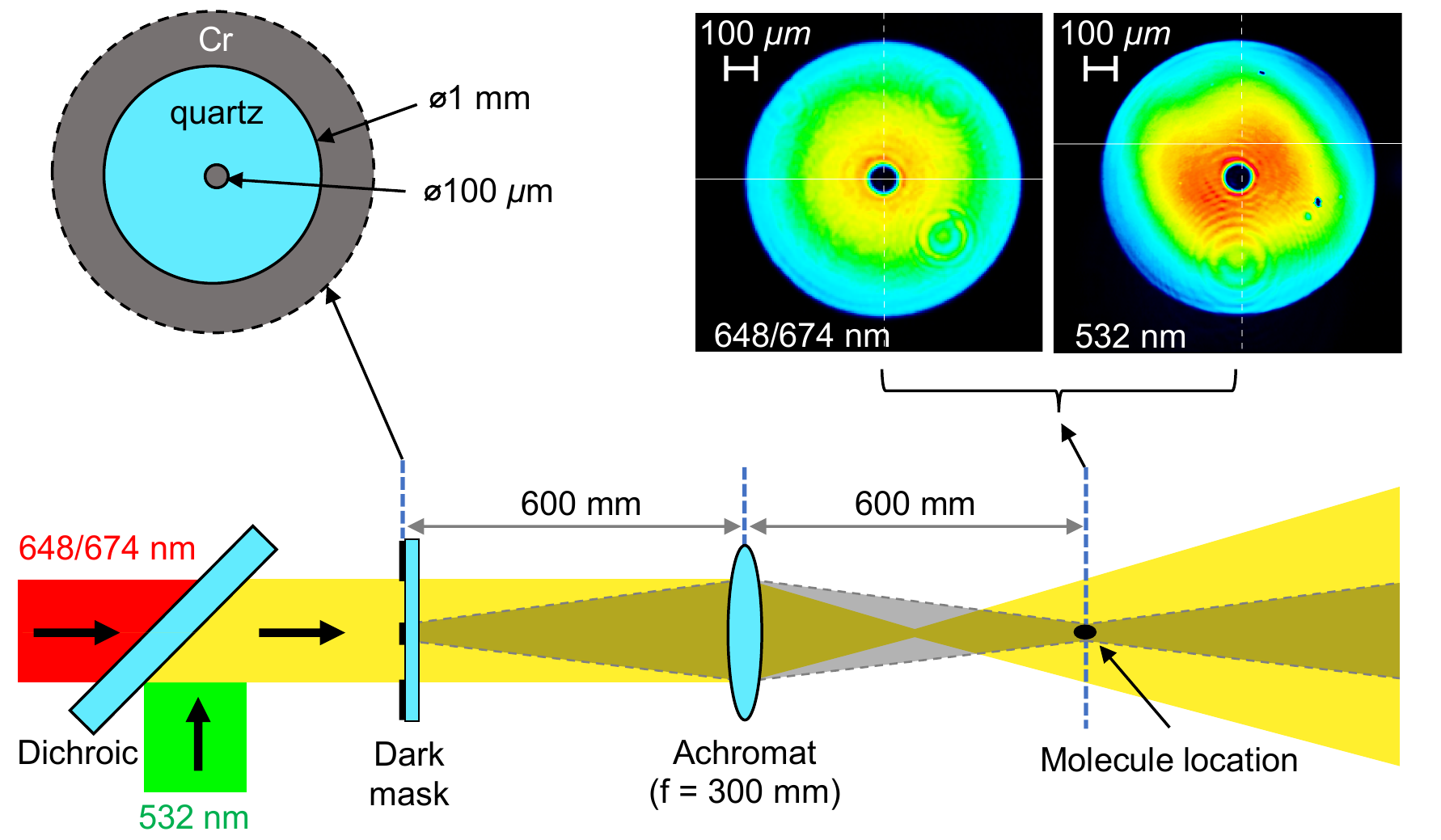}
        \caption{\textbf{Extended Data. The optical setup for generating the REMPI beams.} The 648/674 nm and 532 nm lasers are combined on a dichroic mirror, and are then sent through a dark mask (Thorlabs R1DF100) and an achromatic lens ($f=300$~mm). The resulting beam profiles in the image plane have a 1~mm outer diameter and a $100~\mu$m inner diameter.}
\label{figS1}
\end{figure}

\clearpage
\begin{figure}
\centering
\includegraphics[width=6 in]{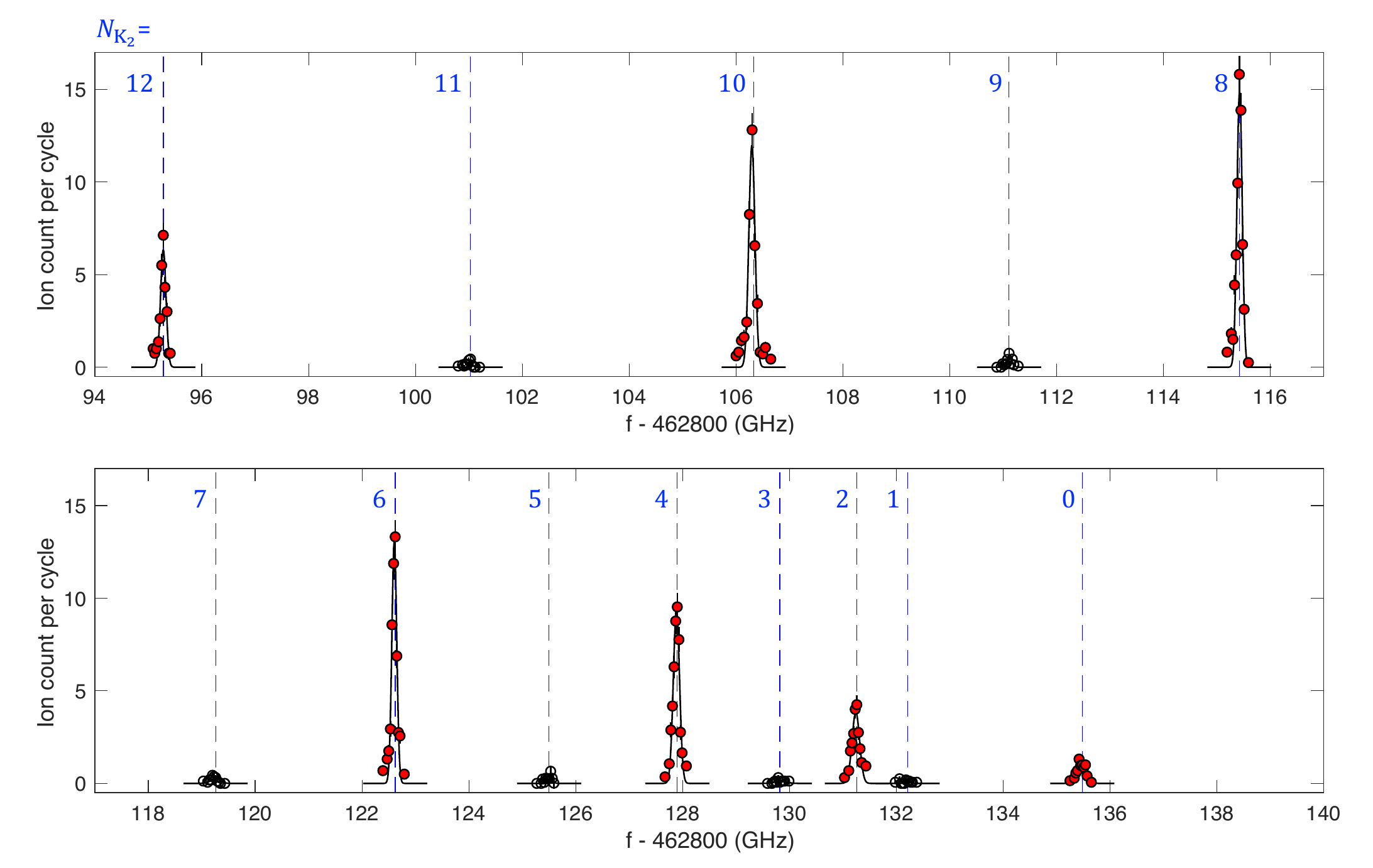}
\caption{\textbf{Extended Data. REMPI spectrum for $^{40}$K$_2$ product molecules at 30 G.} 
By scanning the 648 nm laser frequency to search for rotational lines within the $X^1\Sigma_g^+(v=0,N_{\text{K}_2})\rightarrow B^1\Pi_u(v'=1,N'_{\text{K}_2})$ vibronic band, we observe strong K$_2^+$ signals at frequencies corresponding to rotational states with even values of $N_{\text{K}_2}$ (red filled circles), and highly suppressed signals for odd values (black open circles).
The ion count for each data point is normalized by the corresponding number of experimental cycles ($\sim16$); the error bars denote shot noise.
For $N_{\text{K}_2}>0$ we drive transitions with $N'_{\text{K}_{2}}-N_{\text{K}_{2}}=0$ ($Q$ branch), whereas for $N_{\text{K}_2}=0$, we drive the only allowed transition, with $N'_{\text{K}_{2}}-N_{\text{K}_{2}}=1$ ($R$ branch).
Blue dashed lines indicate the predicted transition frequencies. 
We do not observe any signals at frequencies corresponding to states with $N_{\text{K}_2} > 12$. 
Gaussian fits (black curves) are applied to each signal peak, yielding a typical spectral linewidth (1$\sigma$) of $\sim50$ MHz. 
}
\label{figS2}
\end{figure}

\clearpage

\begin{figure}
\centering
\includegraphics[width=6 in]{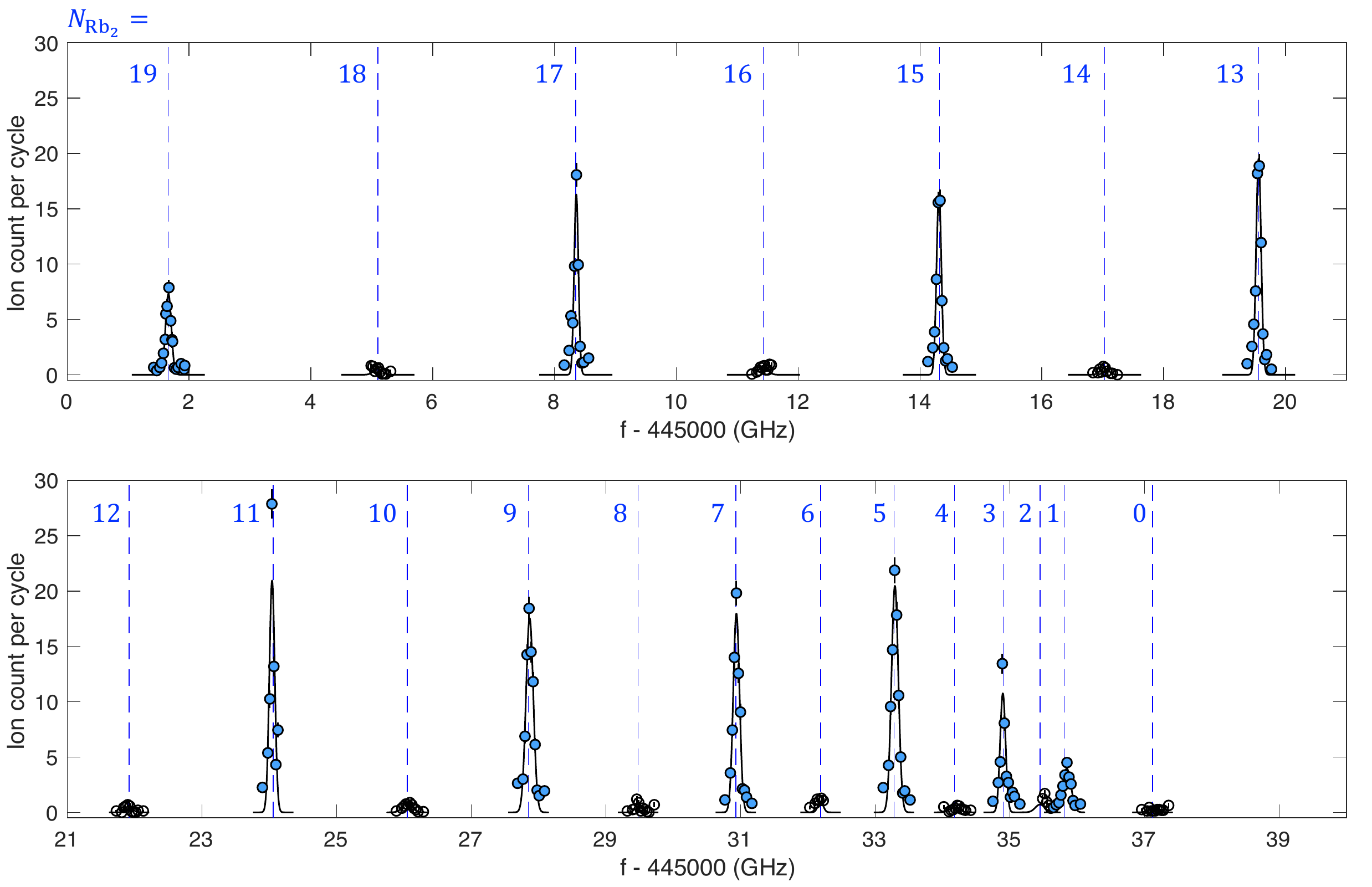}
    \caption{\textbf{Extended Data. REMPI spectrum for $^{87}$Rb$_2$ product molecules at 30 G.} 
    The frequency of the 674 nm laser is scanned within the $X^1\Sigma_g^+(v=0,N_{\text{Rb}_2})\rightarrow B^1\Pi_u(v'=4,N'_{\text{Rb}_2})$ vibronic band. We observe strong Rb$_2^+$ signals for transitions from odd rotational states (blue filled circles), and highly suppressed signals from even ones (black open circles).  The ion count for each data point is normalized by the corresponding number of experimental cycles ($\sim16$); the error bars denote shot noise. We drive $Q$ branch transitions for $N_{\text{Rb}_2}>0$, and $R$ branch for $N_{\text{Rb}_2} = 0$.
    Blue dashed lines indicate the predicted transition frequencies. 
    We do not observe any signals at frequencies corresponding to states with $N_{\text{Rb}_2} > 19$. 
    Gaussian fits (black curves) are applied to each signal peak, yielding a typical spectral linewidth (1$\sigma$) of $\sim40$ MHz. 
    }
\label{figS3}
\end{figure}

\clearpage

\begin{figure}
\centering
\includegraphics[width=6 in]{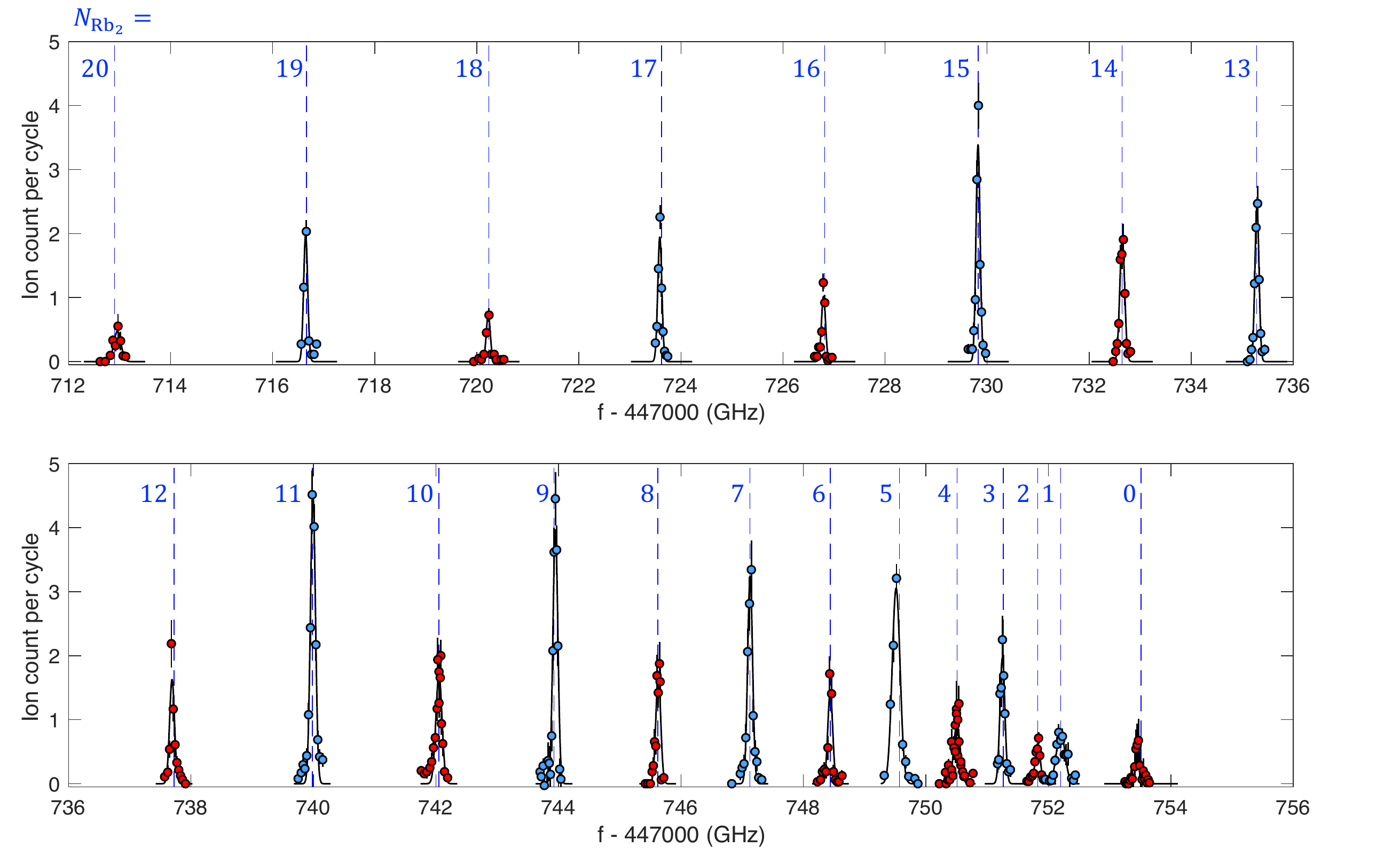}
    \caption{\textbf{Extended Data. REMPI spectrum for $^{87}$Rb$_2$ product molecules at 5 G.} 
    The frequency of the 674 nm laser is scanned within the $X^1\Sigma_g^+(v=0,N_{\text{Rb}_2})\rightarrow B^1\Pi_u(v'=6,N'_{\text{Rb}_2})$ vibronic band. We observe strong Rb$_2^+$ signals for transitions from both even (red filled circles) and odd (blue filled circles) rotational states. The ion count for each data point is normalized by the corresponding number of experimental cycles ($\sim20$); the error bars denote shot noise. We drive $Q$ branch transitions for $N_{\text{Rb}_2}>0$, and $R$ branch for $N_{\text{Rb}_2} = 0$.
    Blue dashed lines indicate the predicted transition frequencies. 
    We do not observe any signals at frequencies corresponding to states with $N_{\text{Rb}_2} > 19$. 
    Gaussian fits (black curves) are applied to each signal peak, yielding a typical spectral linewidth (1$\sigma$) of $\sim40$ MHz. 
    }
\label{figS4}
\end{figure}

\end{document}